\documentclass[11pt]{article}
\usepackage[margin=1in]{geometry}
\usepackage{amsmath,amssymb,amsthm,mathtools,bm}
\usepackage{booktabs,graphicx,caption,enumitem}
\usepackage[numbers,sort&compress]{natbib}
\usepackage{xcolor}
\usepackage[colorlinks=true,linkcolor=blue!50!black,citecolor=blue!50!black,urlcolor=blue!50!black]{hyperref}
\hypersetup{pdftitle={Why the Benjamini-Hochberg procedure tolerates correlated noise, and when it does not},pdfauthor={Lei Sun}}

\newtheorem{theorem}{Theorem}
\newtheorem{proposition}{Proposition}
\newtheorem{lemma}{Lemma}
\newtheorem{corollary}{Corollary}
\theoremstyle{remark}
\newtheorem{remark}{Remark}
\theoremstyle{definition}
\newtheorem{assumption}{Assumption}

\newcommand{\E}{\mathbb{E}}
\renewcommand{\P}{\mathbb{P}}
\newcommand{\Var}{\operatorname{Var}}
\newcommand{\Cov}{\operatorname{Cov}}
\newcommand{\tr}{\operatorname{tr}}
\newcommand{\He}{\mathrm{He}}
\newcommand{\Phib}{\bar\Phi}
\newcommand{\FDP}{\mathrm{FDP}}
\newcommand{\FDR}{\mathrm{FDR}}
\newcommand{\Ghat}{\widehat G}
\newcommand{\Hnull}{\mathcal H_0}
\newcommand{\rhob}{\bar\rho_2}
\newcommand{\tstar}{\tau_\ast}
\newcommand{\cstar}{c_\ast}

\title{Why the Benjamini--Hochberg procedure tolerates correlated noise,\\ and when it does not}
\author{Lei Sun\\[2pt] \normalsize Independent Researcher}
\date{November 2026}

\begin{document}
\maketitle

\begin{abstract}
The Benjamini--Hochberg (BH) procedure is proved to control the false discovery rate (FDR) under
independence and positive regression dependence, yet in genomics-type data with arbitrary
gene--gene correlation and two-sided tests its average false discovery proportion (FDP) is usually
at or below the nominal level, while the FDP of any one data set can be far from it. We explain
both facts through the \emph{realized} distribution of the correlated noise. An exact identity
writes the FDP of BH as $\alpha\pi_0$ times the realized inflation of the null tail at the data-chosen
threshold; since every null $p$-value is uniform, dependence can affect the FDR only through the
coupling between that threshold and the realized null distribution. Under Gaussian noise the
realized null distribution is a smooth random curve indexed by Hermite coefficients, and two-sided
$p$-values see only the even ones. In a large-$p$ factor model we derive the limiting FDP and, in
the regime where a positive fraction of hypotheses is rejected and the correlation is weak, an
expansion in the strength of correlation: the mean FDP moves at order $\rhob$, the mean squared
correlation, whereas the FDP itself fluctuates at order $\rhob^{1/2}$. The leading correction is
the sum of a \emph{co-movement} term, which is negative when the rejections are dominated by nulls,
because false and total discoveries rise and fall together, and a \emph{shift} term, which is
positive and appears only when the dependence moves the non-null statistics coherently. This
anatomy accounts for the recent counterexample of \citet{dobriban2026}, which lies outside the
weak-correlation regime: its excess FDR comes from rare values of the factor that shift the nulls
outward while pulling the signals in. Finally the
identity turns the noise distribution estimated by \texttt{cashr} \citep{sun2018} into an estimate
of the realized FDP of BH on a given data set; on 1{,}000 realistic data sets it has correlation
0.96 with the true FDP.
\end{abstract}

\section{Introduction}

Let $z_j=\theta_j+\varepsilon_j$, $j=1,\dots,p$, with $\varepsilon\sim N_p(0,\Sigma)$ and
$\Sigma_{jj}=1$, and let $p_j=2\Phib(|z_j|)$ be two-sided $p$-values for $H_{0j}:\theta_j=0$. The
BH procedure at level $\alpha$ controls $\FDR=\E[\FDP]$ at $\pi_0\alpha$ when the null $p$-values are
independent \citep{bh1995} or positively regression dependent \citep{by2001,sarkar2002}, but under
arbitrary dependence the guarantee drops to $\alpha\pi_0\sum_{k\le p}k^{-1}$, which is sharp
\citep{by2001,guo2008}. For two-sided Gaussian tests with arbitrary correlation the question of
whether $\FDR\le\alpha$ stood open for decades, supported by simulation and partial results
\citep{reiner2007,benjamini2010}. It has recently been settled in the negative:
\citet{dobriban2026} constructs a one-factor design with $\FDR>0.0104$ at $\alpha=0.01$ for all large
$p$. \citet{lei2026} shows that, as $\alpha\to0$, the worst-case FDR of two-sided BH over Gaussian
correlation matrices is at least $\{1/(2\sqrt\pi)+o(1)\}\,\alpha\sqrt{\log(1/\alpha)}$, so that no
bound of the form $C\alpha$ holds; the same rate is an upper bound within one-factor models, and is
conjectured to be the worst case in general. Further bounds, under correlations that decay with the
distance between indices or in terms of the dependence of each statistic on the others, are given by
\citet{shang2026} and \citet{ghosh2026}. Under the global null, \citet{zhang2026} proves the two-sided
Simes inequality for every number of tests $m$ and every correlation matrix, once $\alpha$ is below a
level that depends on $m$.

These results bound the worst case. Our question is a different one: why, for the correlation found
in real data, is the \emph{average} FDP of BH so stable, when the FDP of each data set is not? Our
starting point is the observation of \citet{efron2007}, \citet{schwartzman2011} and \citet{sun2018}
that correlated null $z$-scores do not follow $N(0,1)$ as an ensemble: their realized empirical
distribution is a random distortion of $N(0,1)$, which can be described by Gaussian derivatives or,
equivalently, Hermite polynomials. Our contributions are as follows.

\begin{enumerate}[leftmargin=*,itemsep=2pt]
\item An exact identity, $\FDP=\alpha\pi_0\,\Ghat_0(\tau)/\tau$, where $\Ghat_0$ is the empirical CDF
  of the null $p$-values and $\tau$ the BH threshold (Lemma~\ref{lem:identity}). Because
  $\E\Ghat_0(t)=t$ for every fixed $t$ under any dependence, $\FDR-\alpha\pi_0$ is entirely a
  coupling effect between $\tau$ and $\Ghat_0$, which Theorem~\ref{thm:cov} writes as a sum of
  covariances given a conditioning factor.
\item The realized null distribution in Hermite form (Proposition~\ref{prop:hermite}), the
  large-$p$ limit of the FDP under a factor model (Theorem~\ref{thm:limit}), and an expansion of the
  limiting FDR to fourth order in the size of the loadings (Theorem~\ref{thm:expansion}). The
  mean shift is $O(\rhob)$ while the standard deviation of the FDP is of order $\rhob^{1/2}$: at a
  root-mean-square correlation of $0.1$ the mean FDP typically moves by about $1\%$ of $\alpha\pi_0$,
  while its standard deviation is about $60\%$ of $\alpha\pi_0$. The expansion is a theorem about
  weak correlation in the dense regime, where the BH threshold stays bounded away from zero; it
  explains the behaviour of BH there and is not a statement of FDR control in general.
\item A sign theorem: in exchangeable designs, the leading correction is negative whenever the
  rejections are dominated by nulls (Corollary~\ref{cor:sign}). A positive correction needs either
  signals that dominate the rejections, in which case the FDR rises above $\pi_0\alpha$ but stays far
  below $\alpha$, or dependence that moves the signals coherently.
\item An anatomy of the counterexample of \citet{dobriban2026} (Section~\ref{sec:dobriban}).
\item A plug-in estimate of the realized FDP of BH from the noise distribution fitted by
  \texttt{cashr} (Section~\ref{sec:plugin}).
\end{enumerate}

The limits we use, in which the empirical $p$-value distribution converges conditionally on a
factor, go back to \citet{genovese2004}, \citet{finner2007} and \citet{delattre2011}; \citet{fan2012}
approximate the FDP under arbitrary covariance by conditioning on principal factors, and
\citet{clarke2009} study the robustness against dependence of procedures designed for independent
tests. The leave-one-out identity behind Theorem~\ref{thm:cov} is standard \citep{ferreira2006};
\citet{lei2026} conditions on the factor in the same way to bound the FDR in one-factor models.

\section{An exact identity}\label{sec:identity}

Let $\Hnull=\{j:\theta_j=0\}$, $p_0=|\Hnull|$, $\pi_0=p_0/p$, and define the empirical CDFs
\[
\Ghat(t)=\frac1p\#\{j:p_j\le t\},\qquad \Ghat_0(t)=\frac1{p_0}\#\{j\in\Hnull:p_j\le t\}.
\]
BH rejects the $R=\max\{k:p_{(k)}\le\alpha k/p\}$ smallest $p$-values ($R=0$ if the set is empty). Write
$\tau=\alpha R/p$ and $V$ for the number of rejected nulls, and set $0/0=0$.

\begin{lemma}\label{lem:identity}
For any joint distribution of $(p_1,\dots,p_p)$:
\begin{enumerate}[label=(\alph*),itemsep=1pt]
\item BH rejects exactly $\{j:p_j\le\tau\}$, so $\Ghat(\tau)=\tau/\alpha$; moreover
  $\tau=\sup\{t\in[0,1]:\alpha\Ghat(t)\ge t\}$.
\item $\FDP=V/R=\alpha\pi_0\,\Ghat_0(\tau)/\tau$.
\item If each null $p$-value is uniform, then $\E\Ghat_0(t)=t$ for every fixed $t$, and
  \[\FDR-\alpha\pi_0=\alpha\pi_0\,\E\big[\Ghat_0(\tau)/\tau-1\big].\]
\end{enumerate}
\end{lemma}

The proof is in Appendix~\ref{app:proofs}. Write $\widehat L(t)=\Ghat_0(t)/t$ for the realized
inflation of the null tail at level $t$. By (b), the FDP is $\alpha\pi_0$ times $\widehat L$ at the
threshold the data choose; by (c), the inflation has mean one at every \emph{fixed} level. Under
independence $t\mapsto\widehat L(t)$ is a backward martingale and $\tau$ a stopping time, so
$\E\widehat L(\tau)=1$ \citep{storey2004}. Under dependence the martingale property fails, and the
FDR departs from $\alpha\pi_0$ only through the dependence between $\tau$ and the curve
$\widehat L$. The next result makes this precise.

\begin{theorem}[Covariance formula]\label{thm:cov}
Let $F$ be a random element such that $p_1,\dots,p_p$ are conditionally independent given $F$, and each
null $p$-value is uniform. Let $G_j(t\mid F)=\P(p_j\le t\mid F)$, $L_j(t\mid F)=G_j(t\mid F)/t$, and
let $R^{(j)}$ be the number of BH rejections when $p_j$ is replaced by $0$. Then
\begin{align}
\FDR&=\frac{\alpha}{p}\sum_{j\in\Hnull}\E\Big[L_j\big(\alpha R^{(j)}/p\;\big|\;F\big)\Big],\label{eq:cov1}\\
\FDR-\alpha\pi_0&=\frac{\alpha}{p}\sum_{j\in\Hnull}\sum_{k=1}^{p}
\Cov\Big(L_j(\alpha k/p\mid F),\ \P(R^{(j)}=k\mid F)\Big).\label{eq:cov2}
\end{align}
\end{theorem}

For Gaussian noise with $\Sigma\succ0$, such an $F$ always exists: with $\delta=\lambda_{\min}(\Sigma)$ and
$BB^\top=\Sigma-\delta I$, take $\varepsilon=BF+\sqrt\delta\,e$ with $F$ and $e$ independent standard
normal vectors. In a factor model, $F$ is the vector of factors. Since
$\E_F L_j(t\mid F)=1$ for each $t$, \eqref{eq:cov2} says that the FDR is inflated exactly when the factor
values that inflate the null tail at level $\alpha k/p$ also make $k$ a likely number of
rejections. Under independence $F$ is trivial and $\FDR=\alpha\pi_0$.

\section{The realized null distribution in Hermite form}\label{sec:hermite}

Let $\He_l$ be the probabilists' Hermite polynomials, orthogonal under $\phi$ with
$\int\He_l\He_m\phi=l!\,\delta_{lm}$, and let $c_t=\Phib^{-1}(t/2)$, so that $\P(|Z|\ge c_t)=t$.

\begin{lemma}\label{lem:psi}
(a) (Mehler) If $(X,Y)$ is standard bivariate normal with correlation $\rho$, then
\[\E[\He_l(X)\He_m(Y)]=\delta_{lm}\,l!\,\rho^l.\]
(b) For $l\ge1$ and $c>0$,
\[
\psi_l(c;\theta):=\int_{\{|\theta+z|\ge c\}}\He_l(z)\phi(z)\,dz
=\phi(c-\theta)\He_{l-1}(c-\theta)+(-1)^l\phi(c+\theta)\He_{l-1}(c+\theta).
\]
In particular $\psi_l(c;0)=2\phi(c)\He_{l-1}(c)$ for even $l$ and $\psi_l(c;0)=0$ for odd $l$.
\end{lemma}

\begin{proposition}\label{prop:hermite}
Let $W_l=p_0^{-1}\sum_{j\in\Hnull}\He_l(\varepsilon_j)$. For every $t\in(0,1)$, in $L^2(\P)$,
\begin{equation}\label{eq:hermite}
\Ghat_0(t)=t+\sum_{l\ge2\ \mathrm{even}}\frac{W_l}{l!}\,\psi_l(c_t;0),\qquad
\widehat L(t)=1+\sum_{l\ge2\ \mathrm{even}}\frac{W_l}{l!}\,\ell_l(t),\quad \ell_l(t)=\frac{\psi_l(c_t;0)}{t},
\end{equation}
with $\E W_l=0$ and $\Cov(W_l,W_m)=\delta_{lm}\,l!\,p_0^{-2}\sum_{i,j\in\Hnull}\rho_{ij}^l$. Hence
\[
\Var\Ghat_0(t)=\sum_{l\ \mathrm{even}}\frac{\psi_l(c_t;0)^2}{l!\,p_0^2}\sum_{i,j\in\Hnull}\rho_{ij}^l .
\]
\end{proposition}

Three features of \eqref{eq:hermite} drive the rest of the paper. First, two-sided $p$-values see
only the even, symmetric modes of the noise: a common shift of the null $z$-scores (the mode
$W_1$, the sample mean of the noise) does not change $\Ghat_0$ to first order. Second, the leading
mode is the dispersion variate $W_2=p_0^{-1}\sum_{j}(\varepsilon_j^2-1)$ of \citet{efron2007}, with
$\Var W_2\approx2\rhob$, where $\rhob$ is the mean squared correlation among nulls; the higher modes
have variances $l!\,\bar\rho_l$ with $\bar\rho_l\le\rhob$, and they are damped by $1/l!$. Third, the
profiles are explicit and smooth. For the dispersion mode,
\[
\ell_2(t)=\frac{2c_t\phi(c_t)}{t}=c_t\,\lambda(c_t),\qquad \lambda(c)=\frac{\phi(c)}{\Phib(c)},
\]
which is strictly decreasing and convex in $t$ (Lemma~\ref{lem:shape}), and behaves like
$c_t^2\approx2\log(1/t)$ as $t\to0$. Overdispersed noise ($W_2>0$) inflates the null tail by a factor
that grows only logarithmically as the threshold moves outwards. The expansion behind
Proposition~\ref{prop:hermite} also underlies the variance calculations of \citet{schwartzman2011};
here we use it as an identity for the curve itself.

\section{The large-\texorpdfstring{$p$}{p} limit under a factor model}\label{sec:limit}

\begin{assumption}\label{ass:factor}
(A1) $\varepsilon_j=s\,b_j^\top F+(1-s^2|b_j|^2)^{1/2}e_j$, with $F\sim N_K(0,I)$ and
$e_1,e_2,\dots$ i.i.d.\ $N(0,1)$, independent of $F$, and $|b_j|\le B$ for all $j$, where $s\ge0$ and
$sB<1$. Thus, given $F$, the $\varepsilon_j$ are independent with variances
$1-s^2|b_j|^2\ge v_0:=1-s^2B^2>0$.
(A2) As $p\to\infty$, $p_0/p\to\pi_0\in(0,1]$, and the empirical distribution of $\{b_j:j\in\Hnull\}$ and
that of $\{(\theta_j,b_j):j\notin\Hnull\}$ converge weakly to $\nu_0$ and $\nu_1$ respectively
(necessarily supported in $\{|b|\le B\}$).
\end{assumption}

Let $Q(c;\mu,v)=\P(|\mu+\sqrt v\,Z|\ge c)$, $\pi_1=1-\pi_0$, and for $f\in\mathbb R^K$
\begin{align*}
G_0(t;f)&=\int Q\big(c_t;\ s\,b^\top f,\ 1-s^2|b|^2\big)\,\nu_0(db),\\
G(t;f)&=\pi_0G_0(t;f)+\pi_1\int Q\big(c_t;\ \theta+s\,b^\top f,\ 1-s^2|b|^2\big)\,\nu_1(d\theta,db),\\
\tau(f)&=\sup\{t\in[0,1]:\alpha G(t;f)\ge t\}.
\end{align*}

\begin{assumption}\label{ass:crossing}
(A3) For almost every $f$: $\tau(f)>0$, and for every $\epsilon>0$ there is
$t\in(\tau(f)-\epsilon,\tau(f)]$ with $\alpha G(t;f)>t$.
\end{assumption}

\begin{theorem}\label{thm:limit}
Under (A1)--(A3), conditionally on $F=f$ and for almost every $f$: $\sup_t|\Ghat(t)-G(t;f)|\to0$ and
$\sup_t|\Ghat_0(t)-G_0(t;f)|\to0$ almost surely, $\tau\to\tau(f)$, and
\[
\FDP\ \to\ \FDP_\infty(f):=\alpha\pi_0\,\frac{G_0(\tau(f);f)}{\tau(f)}.
\]
Consequently $\FDR\to\FDR_\infty:=\E\,\FDP_\infty(F)$.
\end{theorem}

The limit $\FDP_\infty$ is random through the factor, and $\E_F G_0(t;F)=t$ for every $t$, so
$\FDR_\infty-\alpha\pi_0=\alpha\pi_0\,\E[L(\tau(F);F)-1]$ with $L(t;f)=G_0(t;f)/t$. Expanding $Q$ in
Hermite polynomials and using $\E[\He_l(\rho X+\sqrt{1-\rho^2}Z)\mid X]=\rho^l\He_l(X)$ gives
\[
G_0(t;f)=t+\sum_{l\ \mathrm{even}}\frac{W_l(f)}{l!}\,\psi_l(c_t;0),\qquad
W_l(f)=s^l\!\int|b|^l\He_l\big(b^\top f/|b|\big)\,\nu_0(db),
\]
which is the population form of \eqref{eq:hermite}: the realized null distribution in the limit
is the deterministic Hermite expansion with coefficients that are polynomials in the factor.

\begin{remark}[Astronomically small thresholds]\label{rem:tiny}
$\tau(f)$ can be extremely small. Given a large factor value, the null $z$-scores are
$N(\mu,\sigma^2)$ with $\sigma<1$ and $\mu\neq0$, whose tails are heavier than those of $N(0,1)$ up
to about $c=|\mu|/(1-\sigma)$, so $\alpha G(t;f)\ge t$ can hold at very small $t$; in the example
of Section~\ref{sec:dobriban}, $\tau(3)\approx10^{-17}$. Such roots contribute
to $\FDR_\infty$ but are irrelevant for any realistic $p$, since $R\ge1$ requires $\tau\ge\alpha/p$.
For finite $p$ we therefore also use the truncated limit in which the supremum defining $\tau(f)$
is restricted to $t\ge\alpha/p$; neither version describes the sparse regime $R=o(p)$, where the
extremes of the noise, not its bulk, decide the FDP.
\end{remark}

\section{Small-correlation expansion}\label{sec:expansion}

\paragraph{Scope.} This section concerns weak correlation in the dense regime. We fix
$\nu_0,\nu_1$ and let the loading scale $s\to0$, so that every correlation is $O(s^2)$, and we assume
that without correlation BH rejects a positive fraction of the hypotheses. The results describe how
BH behaves there; they are not statements of FDR control for arbitrary correlation, which fails in
general (Section~\ref{sec:dobriban}), and they do not cover the sparse regime, in which the BH
threshold tends to zero.

At $s=0$ the noise is independent and
$G(t)=\pi_0t+\pi_1\int Q(c_t;\theta,1)\,\nu_1(d\theta,db)$. Assume that
$\tstar=\sup\{t:\alpha G(t)\ge t\}\in(0,\alpha]$ and that the crossing is transversal, $\alpha
G'(\tstar)<1$. Put $\cstar=c_{\tstar}$, $\kappa=\alpha/(1-\alpha G'(\tstar))>0$, and
\begin{align*}
M_0&=\int bb^\top\nu_0(db),&
a(t)&=\pi_1\int\psi_1(c_t;\theta)\,b\;\nu_1(d\theta,db),\\
N(t)&=\pi_0\,\psi_2(c_t;0)\,M_0+\pi_1\int\psi_2(c_t;\theta)\,bb^\top\nu_1(d\theta,db).
\end{align*}
Here $a(t)\in\mathbb R^K$ is the first-order response of the rejection rate $G(t;f)$ to the
factor (a coherent shift of the non-null statistics), and $N(t)$ is the second-order response,
through the dispersion of the nulls and the non-nulls. The mean squared correlation among nulls is
$\rhob=s^4\tr(M_0^2)$.

\begin{lemma}\label{lem:shape}
$\ell_2$ is strictly decreasing and strictly convex on $(0,1)$, and $G$ is concave on $(0,1)$.
\end{lemma}

For each $s$, let $\FDR_\infty(s)=\E\,\FDP_\infty(F)$, with $\FDP_\infty$ as in
Theorem~\ref{thm:limit} and $\FDP_\infty(f)=0$ if $\tau(f)=0$.

\begin{theorem}\label{thm:expansion}
Assume (A1)--(A2), $\tstar\in(0,\alpha]$ and $\alpha G'(\tstar)<1$. As $s\to0$, with all functions
evaluated at $\tstar$ and primes denoting $d/dt$,
\begin{equation}\label{eq:expansion}
\frac{\FDR_\infty(s)}{\alpha\pi_0}=1+\frac{s^4}{2}\Big\{\underbrace{\kappa\,\ell_2'\,\tr(M_0N)}_{\text{co-movement}}
+\underbrace{\kappa^2\big(\kappa G''\ell_2'+\ell_2''\big)\,a^\top M_0a+2\kappa^2\ell_2'\,a^\top M_0a'}_{\text{shift}}\Big\}+O(s^6),
\end{equation}
and
\begin{equation}\label{eq:sd}
\operatorname{sd}\big(\FDP_\infty(F)\big)=\alpha\pi_0\,\ell_2(\tstar)\sqrt{\rhob/2}\;\big(1+O(s^2)\big).
\end{equation}
Moreover there is $c>0$ such that, without assuming (A3), $\liminf_{p}\FDR$ and $\limsup_{p}\FDR$
both lie within $O(e^{-c/s^2})$ of $\FDR_\infty(s)$, so \eqref{eq:expansion} holds for them as well.
\end{theorem}

Equations \eqref{eq:expansion} and \eqref{eq:sd} are the quantitative answer to our question. The
mean moves by $O(s^4)=O(\rhob)$; the FDP fluctuates by $O(s^2)=O(\rhob^{1/2})$. The first-order
effect of the realized distortion on the mean cancels exactly, because $L(t;0)\equiv1$ is flat in $t$
and $W_2$ has mean zero; what remains is the second-order coupling between the threshold and the
distortion. With $\ell_2(\tstar)\approx\cstar^2\approx6$--$10$ at typical thresholds, a
root-mean-square correlation of $0.1$ ($\rhob=0.01$) gives a standard deviation of the FDP of about
$40$--$70\%$ of $\alpha\pi_0$, against a mean shift of $0.7$--$2\%$ for the coefficients $C$ we computed
(Tables~\ref{tab:expansion} and~\ref{tab:mc}, and Section~\ref{sec:plugin}).

The two terms of \eqref{eq:expansion} have transparent mechanisms. \emph{Co-movement}: a factor value
that inflates the null tail ($W_2>0$) also raises the number of rejections when $N\succ0$, which moves
the threshold outwards to larger $t$, where the inflation $\ell_2$ is smaller. False and total
discoveries rise together and their ratio corrects itself, so the term is negative when
$\tr(M_0N)>0$. \emph{Shift}: when the factor moves the non-null statistics coherently ($a\ne0$), the
threshold moves at first order, $O(s)$, in a direction unrelated to the sign of the null
distortion, while the size of the distortion grows with $|a^\top F|$. Because $\ell_2$ is convex, the
inflation gained when the threshold moves inwards exceeds that lost when it moves outwards. By
Lemma~\ref{lem:shape}, both parts of the first shift term are nonnegative, and they are positive
when $a\ne0$.

\begin{corollary}[Exchangeable designs]\label{cor:sign}
Suppose $\nu_1=g_1\otimes\nu_0$, that is, the loadings of the non-nulls have the same law as those of
the nulls and are independent of $\theta$, and that either $g_1$ is symmetric about $0$ or
$\int b\,\nu_0(db)=0$. Then $a\equiv0$, $N(t)=\Psi_2(t)M_0$ with
$\Psi_2(t)=\pi_0\psi_2(c_t;0)+\pi_1\int\psi_2(c_t;\theta)\,g_1(d\theta)$, and
\begin{equation}\label{eq:exch}
\FDR_\infty=\alpha\pi_0\Big[1+\tfrac12\,\kappa\,\ell_2'(\tstar)\,\Psi_2(\tstar)\,\rhob\Big]+O(s^6).
\end{equation}
Since $\ell_2'<0$, $\FDR_\infty<\alpha\pi_0$ for all small $s>0$ if $\Psi_2(\tstar)>0$, and
$\FDR_\infty>\alpha\pi_0$ if $\Psi_2(\tstar)<0$. The first case holds in particular if (i) $g_1$ is supported in $[-\cstar,\cstar]$, or (ii)
$\cstar\lambda(\cstar)>2\phi(1)\big(1/(\alpha\pi_0)-1\big)$.
\end{corollary}

Condition (ii) is crude: it holds for $\alpha=0.1$, $\pi_0=0.9$ once $\tstar\le0.036$, but
$\Psi_2(\tstar)>0$ in every exchangeable design we tried except the one in which the
rejections are dominated by strong signals (design C in Table~\ref{tab:expansion}). There,
$\FDR_\infty$ exceeds $\pi_0\alpha$ by a relative amount $C\rhob$ with $C=0.20$, while remaining
about ten times below $\alpha$. In exchangeable designs, then, correlation cannot make BH
anti-conservative relative to $\alpha$ at the leading order.

\begin{remark}[One-sided tests]
For one-sided $p$-values the null curve also contains the shift mode, since
$\psi_1(c;0)=\phi(c)\ne0$ in the one-sided analogue of Lemma~\ref{lem:psi}. The leading correction is
then of order $s^2$ rather than $s^4$, and for nonnegative loadings it is nonpositive by the PRDS
theorem of \citet{by2001}. We do not pursue this case.
\end{remark}

\subsection{Numerical checks}

Table~\ref{tab:expansion} compares $\FDR_\infty(s)$, computed by solving the fixed point
$t=\alpha G(t;f)$ on a fine grid in $f$, with \eqref{eq:expansion} in five one-factor designs. In every
design the two agree to within $3\%$ at the smallest $s$ shown. In design D, an aligned variant of A
in which all signals have the same sign and loading, the shift term reverses the sign of A's
correction at small $s$, but higher-order terms take over by $s=0.3$. Table~\ref{tab:mc} checks \eqref{eq:exch} and
\eqref{eq:sd} by simulating BH with $p=10{,}000$ under a five-factor model.

\begin{table}[t]
\centering\small
\caption{Relative FDR excess $10^3\,[\FDR_\infty/(\alpha\pi_0)-1]$: exact large-$p$ limit versus the
fourth-order term $Cs^4$ of \eqref{eq:expansion}. Designs: one factor; blocks
(proportion, $\theta$, loading). A: $(0.9,0,1),(0.05,\pm3,1)$; B: $(0.8,0,1),(0.1,\pm2.5,1)$; C:
$(0.1,0,1),(0.45,\pm3,1)$; D (aligned): $(0.9,0,1),(0.1,3,1)$; all with $\alpha=0.1$. E is
\citet{dobriban2026}: $(0.96,0,0.3),(0.01,2.4,-0.3),(0.03,4.4,-0.72)$, $\alpha=0.01$, $s$ scaling
all loadings.}
\label{tab:expansion}
\begin{tabular}{llrrrrr}
\toprule
design & & $s=0.1$ & $0.2$ & $0.3$ & $0.4$ & $0.5$ \\
\midrule
A ($C=-0.721$) & limit & $-0.072$ & $-1.17$ & $-6.13$ & $-20.5$ & $-52.5$ \\
 & $Cs^4$ & $-0.072$ & $-1.15$ & $-5.84$ & $-18.5$ & $-45.0$ \\[2pt]
B ($C=-0.893$) & limit & $-0.089$ & $-1.39$ & $-6.81$ & $-20.7$ & $-48.3$ \\
 & $Cs^4$ & $-0.089$ & $-1.43$ & $-7.24$ & $-22.9$ & $-55.8$ \\[2pt]
C ($C=+0.204$) & limit & $+0.020$ & $+0.33$ & $+1.71$ & $+5.6$ & $+13.9$ \\
 & $Cs^4$ & $+0.020$ & $+0.33$ & $+1.65$ & $+5.2$ & $+12.7$ \\[2pt]
D ($C=+0.104$) & limit & $+0.011$ & $+0.15$ & $+0.05$ & $-5.7$ & $-34.9$ \\
 & $Cs^4$ & $+0.010$ & $+0.17$ & $+0.84$ & $+2.6$ & $+6.5$ \\
\midrule
 & & $s=0.2$ & $0.4$ & $0.6$ & $0.8$ & $1.0$ \\
\midrule
E ($C=+0.026$) & limit & $+0.042$ & $+0.75$ & $+4.77$ & $+32.1$ & $+210.6$ \\
 & $Cs^4$ & $+0.041$ & $+0.66$ & $+3.33$ & $+10.5$ & $+25.7$ \\
\bottomrule
\end{tabular}
\end{table}

\begin{table}[t]
\centering\small
\caption{BH with $p=10{,}000$, $\pi_0=0.9$, $\theta=\pm3$ with random signs, $\alpha=0.1$, under a
five-factor model with loadings uniform on a sphere and independent of $\theta$ (4{,}000 replicates
for each $\rhob$; simulated means are $\pm$ one standard error). Predictions are from
\eqref{eq:exch} and \eqref{eq:sd}; the predicted standard deviation adds the independent-noise value
$0.126$ in quadrature.}
\label{tab:mc}
\begin{tabular}{ccccc}
\toprule
& \multicolumn{2}{c}{mean FDP$/(\alpha\pi_0)$} & \multicolumn{2}{c}{sd FDP$/(\alpha\pi_0)$} \\
\cmidrule(lr){2-3}\cmidrule(l){4-5}
$\rhob$ (rms correlation) & predicted & simulated & predicted & simulated \\
\midrule
0 & 1 & $0.999\pm0.002$ & -- & 0.126 \\
0.005 (0.07) & 0.996 & $0.993\pm0.007$ & 0.43 & 0.42 \\
0.010 (0.10) & 0.993 & $0.998\pm0.009$ & 0.59 & 0.58 \\
0.020 (0.14) & 0.986 & $0.993\pm0.013$ & 0.83 & 0.81 \\
\bottomrule
\end{tabular}
\end{table}

\section{Anatomy of a counterexample}\label{sec:dobriban}

The design of \citet{dobriban2026} is design E with $s=1$: one factor, $\alpha=0.01$, nulls with loading
$0.3$, and two signal blocks with means $2.4$ and $4.4$ and loadings $-0.3$ and $-0.72$. Our
computation of the limit gives $\FDR_\infty=0.01162$, consistent with his certified lower bound
$0.0104$. At finite $p$ the convergence is slow: our conditional Monte Carlo gives $0.01019$
(standard error $0.00006$) at $p=20{,}000$, close to the value $0.01036$ (standard error $0.00010$)
reported by \citet{dobriban2026}.

\begin{figure}[t]
\centering
\includegraphics[width=0.85\textwidth]{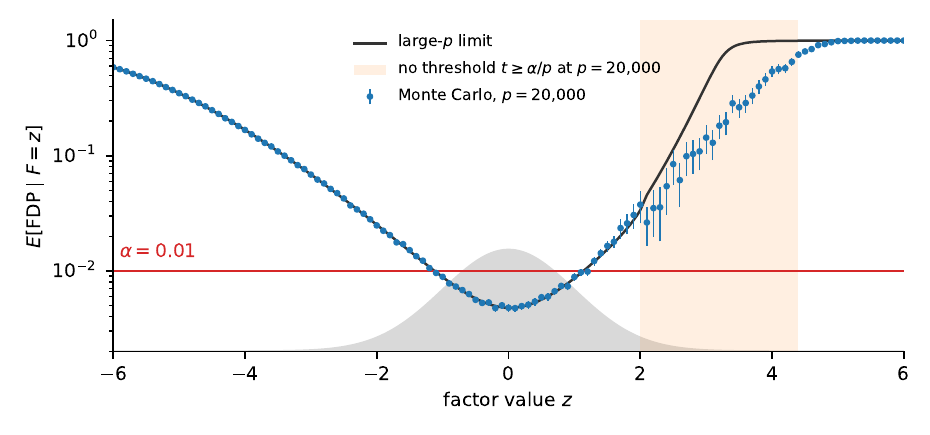}
\caption{Design of \citet{dobriban2026}: conditional FDR given the factor value $z$, from Monte Carlo
($p=20{,}000$, 300 replicates per value, $\pm2$ standard errors) and the large-$p$ limit, with a
shaded band where no threshold $t\ge\alpha/p$ exists at $p=20{,}000$. The grey curve is the density
of the factor.}
\label{fig:dobriban}
\end{figure}

Figure~\ref{fig:dobriban} shows the conditional FDR $\E[\FDP\mid F=z]$, and Table~\ref{tab:dobriban}
splits the FDR by the value of the factor. For the $94\%$ of factor values below $1.5$, BH is
conservative: these values contribute $0.0078$ against the nominal $\alpha\pi_0\P(F<1.5)=0.0090$, and
the large-$p$ limit agrees with the simulation to within $10^{-5}$. The violation comes entirely from the
$6\%$ of factor values above $1.5$. There the nulls shift outwards (mean $0.3z$), while both signal
blocks are pulled towards zero (means $2.4-0.3z$ and $4.4-0.72z$). The threshold collapses, to
$\tau(3)\approx10^{-17}$ in the limit, and almost every rejection is a null: in the limit
$\E[\FDP\mid F=z]$ exceeds $0.9$ for $z\ge3.5$. At $p=20{,}000$ no threshold $t\ge\alpha/p$ crosses
for $2\le z\le4.4$, the rejections there come from the extremes of the noise, and the conditional
FDR rises more slowly, reaching $0.5$ at $z=4$. This is the shift mechanism of Section~\ref{sec:expansion} in extreme
form: the dependence moves signals and nulls in opposite directions. It is not a small-correlation
effect, since at $s=1$ the exact excess, $+0.21$, is eight times the fourth-order term.

\begin{table}[t]
\centering\small
\caption{Design of \citet{dobriban2026}: contributions to the FDR by the value of the factor.}
\label{tab:dobriban}
\begin{tabular}{lcccc}
\toprule
factor region & probability & Monte Carlo, $p=20{,}000$ & limit $p=\infty$ & $\alpha\pi_0\times$probability \\
\midrule
$F<1.5$ & 0.936 & 0.00783 & 0.00784 & 0.00899 \\
$F\ge1.5$ & 0.064 & 0.00236 & 0.00378 & 0.00061 \\
total & 1 & 0.01019 & 0.01162 & 0.00960 \\
\bottomrule
\end{tabular}
\end{table}

\section{Estimating the realized FDP of BH with cashr}\label{sec:plugin}

Lemma~\ref{lem:identity}(b) says that the FDP of a BH run is $\alpha\pi_0\widehat L(\tau)$, and $\tau$
is observed. What is unknown is the realized null tail $\Ghat_0(\tau)$, which is exactly what the
correlated-noise model of \texttt{cashr} \citep{sun2018} estimates.\footnote{We use version 1.0.0 of
the implementations in R (\url{https://github.com/LSun/cashr}) and Python
(\url{https://github.com/LSun/cashpy}), which give identical fits.} Its fitted noise density is
$f(z;\omega)=\sum_{l=0}^L\omega_l\phi^{(l)}(z)/\sqrt{l!}$, shared by nulls and non-nulls, with tail
\[
T(t;\omega)=\int_{|z|\ge c_t}f(z;\omega)\,dz=t+2\phi(c_t)\sum_{l\ge2\ \mathrm{even}}\omega_l\,
\frac{h_{l-1}(c_t)}{\sqrt l},\qquad h_l=\He_l/\sqrt{l!},
\]
by Lemma~\ref{lem:psi}. With $\widehat\pi_0$ and $\widehat\omega$ from \texttt{cashr}, the plug-in
estimate of the realized FDP of BH is
\[
\widehat{\FDP}=\alpha\,\widehat\pi_0\,T(\tau;\widehat\omega)/\tau .
\]
It estimates the FDP of the data set at hand, not the FDR; it is the BH analogue of the conditional
FDR of \citet{efron2007} and the FDP approximation of \citet{fan2012}.

We applied it to 1{,}000 realistic data sets generated as in Section~4.1 of \citet{sun2018}. In each,
the noise comes from comparing two random groups of five GTEx liver samples \citep{gtex2017} over
10{,}000 genes, so it carries the real gene--gene correlation; $x_j=\theta_j+s_jz_j$ with the
standard errors $s_j$ of the same comparison, $\pi_0=0.9$, and $\theta\sim N(0,2^2)$ for the signals,
assigned to random genes. The
correlation is strong: across data sets $\Var W_2=0.153$, so $\rhob\approx0.077$ (root-mean-square
correlation $0.28$). BH at $\alpha=0.1$ has mean FDP $0.080$, below $\alpha\pi_0=0.09$, and standard
deviation $0.128$. For this exchangeable design \eqref{eq:exch} predicts a mean shift of $-15\%$ and
\eqref{eq:sd} a standard deviation of $2.1\,\alpha\pi_0$, against the observed $-11\%$ and
$1.4\,\alpha\pi_0$. The signs and orders of magnitude are right, although $\rhob$ is well outside
the small-correlation regime and the $s_j$ vary across genes.

\begin{table}[t]
\centering\small
\caption{Estimates of the realized FDP of BH ($\alpha=0.1$) on 1{,}000 realistic data sets. ``Known
noise'' replaces $T(\tau;\widehat\omega)$ by the realized tail fraction of the true noise of all
10{,}000 genes, and $\widehat\pi_0$ by $\pi_0$.}
\label{tab:plugin}
\begin{tabular}{lccc}
\toprule
estimate of FDP & correlation with FDP & RMSE & mean \\
\midrule
nominal $\alpha\widehat\pi_0$ & -- & 0.128 & 0.090 \\
\texttt{cashr} plug-in $\widehat\FDP$ & 0.962 & 0.052 & 0.116 \\
known noise & 0.999 & 0.004 & 0.080 \\
\midrule
true FDP & & & 0.080 \\
\bottomrule
\end{tabular}
\end{table}

\begin{figure}[t]
\centering
\includegraphics[width=0.85\textwidth]{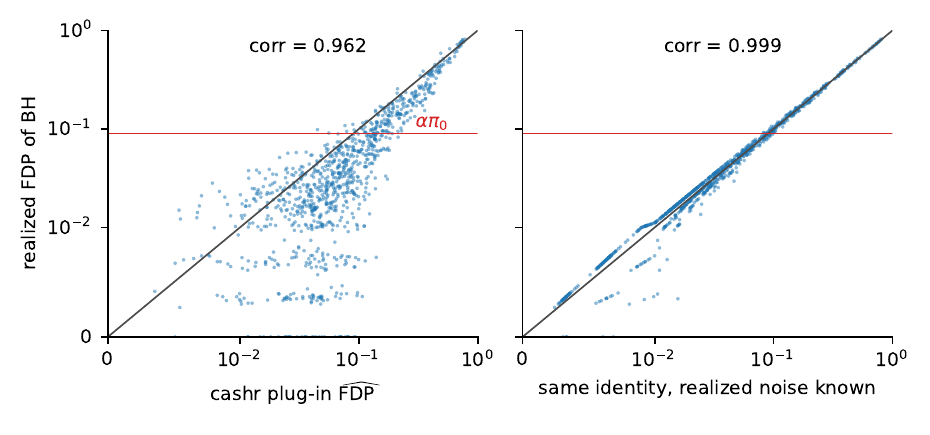}
\caption{Realized FDP of BH on 1{,}000 realistic data sets against the \texttt{cashr} plug-in
estimate (left) and against the same identity with the realized noise known (right). Red line:
nominal $\alpha\pi_0$.}
\label{fig:plugin}
\end{figure}

Table~\ref{tab:plugin} and Figure~\ref{fig:plugin} show the results. The known-noise version is
nearly exact, which confirms both Lemma~\ref{lem:identity} and that nulls and non-nulls share one
realized noise distribution. The \texttt{cashr} estimate tracks the realized FDP with correlation
$0.96$ and cuts the error of the nominal value by more than half. It errs on the high side: by
quintile of $\widehat\FDP$, the mean FDP is $0.015,0.019,0.032,0.062,0.271$ against estimates of
$0.019,0.042,0.068,0.116,0.335$. This is the direction one expects from the conservative default
of \texttt{cashr}, which starts from ``everything is null'' and attributes ambiguous tail mass to
the noise.

For the leukemia data of \citet[Table~1]{sun2018}, 7{,}128 genes, BH at $\alpha=0.1$ makes $1{,}579$
discoveries ($\tau=0.022$), as reported there. \texttt{cashr} estimates $\widehat\pi_0=0.81$ and a
realized null tail inflated by a factor of $6.6$ at $\tau$, so $\widehat\FDP=0.54$. \texttt{cashr}
itself makes 387 discoveries at $q\le0.1$. \citet{sun2018} report 385: their implementation stopped
after at most 50 accelerated EM iterations, often before convergence, whereas version 1.0.0
iterates to convergence and gives 387 whichever block of parameters is updated first.

\section{Discussion}

Our account of why BH is robust has three parts. Marginal uniformity removes, exactly, every effect
of the realized null distortion on the mean FDP except its coupling with the threshold
(Lemma~\ref{lem:identity}, Theorem~\ref{thm:cov}); this holds for every $p$ and every dependence.
Gaussian correlation produces distortions that are smooth, low-dimensional and, for two-sided tests,
symmetric, and their leading mode, the dispersion of the noise, moves false and total discoveries
together. Consequently, for weak correlation in the dense regime, the mean FDP moves at second order,
usually downwards, while the FDP of each data set moves at first order
(Theorem~\ref{thm:expansion}, Corollary~\ref{cor:sign}).

These results explain the typical behaviour of BH; they do not establish FDR control. Within the
expansion, an upward correction needs either rejections dominated by signals, which keeps the FDR
far below $\alpha$, or a factor that moves the signals coherently. Outside it, FDR control can fail:
in the counterexample of Section~\ref{sec:dobriban}, the violation is produced by rare factor values
that move signals and nulls in opposite directions. Whether such alignment occurs in real data is an
empirical question: it would require the signs of the effects to be aligned with co-expression
modules. Simulations that assign signals to random genes, as most do, are exchangeable by
construction.

Open questions:
\begin{enumerate}[leftmargin=*,itemsep=2pt]
\item The sparse regime, $\tstar=0$ or $R=o(p)$, and the extreme factor values of
  Section~\ref{sec:dobriban} lie outside our expansion and call for extreme-value arguments.
\item Is $\FDR\le\alpha$ for exchangeable designs at every correlation strength?
\item A bias-corrected plug-in, and intervals for the realized FDP.
\end{enumerate}

\section*{Acknowledgments}
The author used Claude (Anthropic), a generative AI assistant, to help prepare the draft.
The author reviewed all content and takes full responsibility for it.

\appendix
\section{Proofs}\label{app:proofs}

\paragraph{Lemma~\ref{lem:identity}.}
(a) Let $N(t)=\#\{j:p_j\le t\}$, so that $p_{(k)}\le t$ if and only if $N(t)\ge k$. If $R=0$ there is
nothing to prove. Otherwise $N(\alpha R/p)\ge R$, since $p_{(R)}\le\alpha R/p$. Let $R'=N(\alpha R/p)$.
Then $p_{(R')}\le\alpha R/p\le\alpha R'/p$, so $R'\le R$ by the maximality of $R$, and hence $R'=R$.
The set $\{j:p_j\le\tau\}$ therefore has exactly $R$ elements and consists of the $R$ smallest
$p$-values, and $\Ghat(\tau)=R/p=\tau/\alpha$. For the supremum, let $t\ge\tau$ with $\alpha\Ghat(t)\ge t$
and put $k=N(t)\ge pt/\alpha$. Then $p_{(k)}\le t\le\alpha k/p$, so $k\le R$ and
$t\le\alpha k/p\le\tau$. (b) $V=p_0\Ghat_0(\tau)$ and $R=p\tau/\alpha$. (c) $\E\Ghat_0(t)=p_0^{-1}\sum_{j\in\Hnull}\P(p_j\le t)=t$; then apply (b).

\paragraph{Leave-one-out lemma \citep{ferreira2006}.}
For each $j$: $\{j\text{ rejected}\}=\{p_j\le\alpha R^{(j)}/p\}$, and on this event $R=R^{(j)}$.
\emph{Proof.} Let $N^{(j)}(t)$ count the $p$-values $\le t$ after $p_j$ is set to $0$, so that
$N^{(j)}(t)=N(t)$ for $t\ge p_j$ and $N^{(j)}\ge N$. If $j$ is rejected, then $p_j\le\alpha R/p$, so
$N^{(j)}(\alpha k/p)=N(\alpha k/p)<k$ for every $k>R$ and $N^{(j)}(\alpha R/p)\ge R$; hence
$R^{(j)}=R$ and $p_j\le\alpha R^{(j)}/p$. Conversely, if $p_j\le\alpha R^{(j)}/p=:\alpha k/p$, then
$N(\alpha k/p)=N^{(j)}(\alpha k/p)\ge k$, so $R\ge k$. Also $R^{(j)}\ge R$, because $N^{(j)}\ge N$.
So $R=k$, and $j$ is rejected by (a).

\paragraph{Theorem~\ref{thm:cov}.}
By the lemma, $V/R=\sum_{j\in\Hnull}\mathbf 1\{p_j\le\alpha R^{(j)}/p\}/R^{(j)}$, with $R^{(j)}\ge1$. Given
$F$, $p_j$ is independent of $R^{(j)}$, which is a function of $(p_i)_{i\ne j}$. Hence
$\E[\mathbf 1\{p_j\le\alpha R^{(j)}/p\}/R^{(j)}\mid F]=\E[G_j(\alpha R^{(j)}/p\mid F)/R^{(j)}\mid F]
=(\alpha/p)\,\E[L_j(\alpha R^{(j)}/p\mid F)\mid F]$, which gives \eqref{eq:cov1}. For \eqref{eq:cov2},
write $\E[L_j(\alpha R^{(j)}/p\mid F)]=\sum_k\E[L_j(\alpha k/p\mid F)\,\pi_{jk}(F)]$ with
$\pi_{jk}(F)=\P(R^{(j)}=k\mid F)$, and use $\E L_j(\alpha k/p\mid F)=1$ and $\sum_k\E\pi_{jk}=1$.

\paragraph{Lemma~\ref{lem:psi}.}
(a) is Mehler's formula. (b) From $\phi\He_l=(-1)^l\phi^{(l)}$ we get
$\frac{d}{dz}[\phi\He_{l-1}]=-\phi\He_l$, so $\int_u^\infty\phi\He_l=\phi(u)\He_{l-1}(u)$ and
$\int_{-\infty}^{v}\phi\He_l=-\phi(v)\He_{l-1}(v)$. Apply these with $u=c-\theta$ and $v=-c-\theta$, and
use $\He_{l-1}(-x)=(-1)^{l-1}\He_{l-1}(x)$.

\paragraph{Proposition~\ref{prop:hermite}.}
Fix $t$ and let $u$ be the indicator of $\{|z|\ge c_t\}$. Then $u\in L^2(\phi)$, and its
Hermite coefficients $\E[u(Z)\He_l(Z)]$ are $t$ for $l=0$ and $\psi_l(c_t;0)$ for $l\ge1$. Since
$\varepsilon_j\sim N(0,1)$,
\[u(\varepsilon_j)=t+\sum_{l\ge1}\frac{\psi_l(c_t;0)}{l!}\He_l(\varepsilon_j)\quad\text{in }L^2(\P);\]
average over $j\in\Hnull$ and use Lemma~\ref{lem:psi}(b) to discard the odd $l$. The moments follow from
Lemma~\ref{lem:psi}(a).

\paragraph{Theorem~\ref{thm:limit}.}
Fix $f$. Given $F=f$, the indicators $\mathbf 1\{p_j\le t\}$ are independent with means
$Q(c_t;\theta_j+sb_j^\top f,1-s^2|b_j|^2)$. By Hoeffding's inequality and the Borel--Cantelli lemma,
$\Ghat(t)$ minus its conditional mean tends to $0$ almost surely at each $t$ in a finite grid. By
(A1), the conditional variances satisfy $1-s^2|b|^2\ge v_0>0$ on $\{|b|\le B\}$, which contains the
supports of the empirical distributions and of $\nu_0,\nu_1$. On this set the integrand
$(\theta,b)\mapsto Q(c_t;\theta+sb^\top f,1-s^2|b|^2)$ is bounded by $1$ and continuous, so by
(A2) the conditional mean tends to $G(t;f)$. Monotonicity in $t$ and continuity of
$G(\cdot;f)$ make the convergence uniform, as in the proof of the Glivenko--Cantelli theorem; the
same holds for $\Ghat_0$. For the threshold: on $[\tau(f)+\epsilon,1]$, $\alpha G(t;f)-t\le-\eta<0$,
so eventually $\alpha\Ghat(t)<t$ there and $\tau\le\tau(f)+\epsilon$ by Lemma~\ref{lem:identity}(a).
By (A3) there is $t_0>\tau(f)-\epsilon$ with $\alpha G(t_0;f)>t_0$, so eventually $\alpha\Ghat(t_0)\ge t_0$
and $\tau\ge t_0$. Then $\FDP=\alpha\pi_0\Ghat_0(\tau)/\tau\to\FDP_\infty(f)$ by uniform convergence,
the continuity of $G_0(\cdot;f)$ and $\tau(f)>0$. Since $\FDP\le1$, dominated convergence gives the
FDR.

\paragraph{Lemma~\ref{lem:shape}.}
Let $\lambda(c)=\phi(c)/\Phib(c)$, so that $\lambda'=\lambda(\lambda-c)$ and $\lambda(c)>c$ for all $c$.
Write $\ell_2(t)=h(c_t)$ with $h(c)=c\lambda(c)$, and note $dc_t/dt=-1/(2\phi(c_t))$ and $c_t>0$. Then
$h'(c)=\lambda(1+c\lambda-c^2)$, so
\[
\ell_2'(t)=-\frac{1+c\lambda-c^2}{2\Phib(c)}\Big|_{c=c_t}<0,\qquad
\ell_2''(t)=\frac{(\lambda-c)(1+c\lambda)}{2\phi(c)\Phib(c)}\Big|_{c=c_t}>0,
\]
where the second identity uses
$\frac{d}{dc}\big[(1+c\lambda-c^2)/\Phib\big]=2(\lambda-c)(1+c\lambda)/\Phib$. For $G$: since
$Q(c;\theta,1)=\Phib(c-\theta)+\Phib(c+\theta)$,
$\frac{d}{dt}Q(c_t;\theta,1)=\{\phi(c_t-\theta)+\phi(c_t+\theta)\}/\{2\phi(c_t)\}=e^{-\theta^2/2}\cosh(\theta
c_t)$, which is nonincreasing in $t$. Hence $G'$ is nonincreasing.

\paragraph{Theorem~\ref{thm:expansion}.}
Constants $C,c,r,r'>0$ below depend on $\nu_0,\nu_1,\alpha,B$ but not on $s$ or $f$.

\emph{Step 1 (smooth parametrization).} For $x\in\mathbb R^K$ and $y\in\mathbb R$ let
\[
\mathcal G_0(t;x,y)=\int Q(c_t;b^\top x,1-y|b|^2)\,\nu_0(db),\qquad
\mathcal G=\pi_0\mathcal G_0+\pi_1\!\int Q(c_t;\theta+b^\top x,1-y|b|^2)\,\nu_1(d\theta,db),
\]
and $\Lambda=\mathcal G_0/t$. Then $G_0(t;f)=\mathcal G_0(t;sf,s^2)$, $G(t;f)=\mathcal G(t;sf,s^2)$,
$L(t;f)=\Lambda(t;sf,s^2)$ and $G(t)=\mathcal G(t;0,0)$. Let $U=\{|x|\le r,\ |y|\le r\}$ with
$rB^2\le\frac12$, so that $1-y|b|^2\in[\frac12,\frac32]$. For $v$ in this range, every partial
derivative of $Q(c;\mu,v)=\Phib((c-\mu)/\sqrt v)+\Phib((c+\mu)/\sqrt v)$ in $(c,\mu,v)$ is a finite
sum of terms of the form (polynomial in $w$ and $v^{-1/2}$)$\times\phi(w)$, with
$w=(c\mp\mu)/\sqrt v$, and is therefore bounded uniformly in $c,\mu\in\mathbb R$. Since $|b|\le B$ and
$c_t$ is smooth on $(0,1)$, dominated convergence shows that $\mathcal G_0$, $\mathcal G$ and $\Lambda$
are $C^\infty$ on $(0,1)\times U$, with derivatives bounded on $[t_1,t_2]\times U$ for every
$[t_1,t_2]\subset(0,1)$, and that $|\mathcal G(t;x,y)-\mathcal G(t;0,0)|\le C(|x|+|y|)$ for all
$t\in(0,1]$. Two further facts are used below. $\Lambda$ is even in $x$, because $Q$ is even in $\mu$.
And $\Lambda(t;0,0)=1$.

\emph{Step 2 (the threshold).} Let $D(t;x,y)=\alpha\mathcal G(t;x,y)-t$ and $D(t)=D(t;0,0)$, so that
$D(\tstar)=0$, $D'(\tstar)=-\alpha/\kappa<0$, and $D<0$ on $(\tstar,1]$. Choose $\delta\in(0,\tstar/2)$
with $D'\le-\alpha/(2\kappa)$ on $I=[\tstar-\delta,\tstar+\delta]$. Then $D(\tstar-\delta)>0$, and
$\eta=\min\{D(\tstar-\delta),\,-\max_{[\tstar+\delta,1]}D\}>0$. Shrinking $r$, we may assume that for
$(x,y)\in U$, $|D(t;x,y)-D(t)|<\eta$ for all $t\in(0,1]$, and $\partial_tD(t;x,y)<-\alpha/(4\kappa)$
on $I$. Then, for $(x,y)\in U$, $D(\cdot;x,y)$ is negative on $[\tstar+\delta,1]$, positive at
$\tstar-\delta$, and strictly decreasing on $I$. Consequently its largest zero
$T(x,y)=\sup\{t:D(t;x,y)\ge0\}$ is the unique zero in $I$, and $D(\cdot;x,y)>0$ just to the left of
it. So (A3) holds at every $f$ with $(sf,s^2)\in U$, and $\tau(f)=T(sf,s^2)$. By the implicit
function theorem, $T$ is $C^\infty$ on $U$ with $T(0,0)=\tstar$.

\emph{Step 3 (Taylor expansion).} Let $\Phi(x,y)=\Lambda(T(x,y);x,y)$, which is $C^\infty$ on $U$.
Then $\FDP_\infty(f)=\alpha\pi_0\Phi(sf,s^2)$ whenever $(sf,s^2)\in U$. By Taylor's theorem at order
$5$, $\Phi(x,y)=\sum_{|\beta|+\gamma\le5}c_{\beta\gamma}x^\beta y^\gamma+R(x,y)$ with
$|R(x,y)|\le C(|x|+|y|)^6$ on $U$. Substituting $x=sf$, $y=s^2$ and collecting powers of $s$ gives
\begin{equation}\label{eq:taylor}
\Phi(sf,s^2)=\sum_{k=0}^{5}s^kP_k(f)+\widetilde R_s(f),\qquad |\widetilde R_s(f)|\le Cs^6(1+|f|)^6.
\end{equation}
Here $P_k(f)=\sum_{|\beta|+2\gamma=k}c_{\beta\gamma}f^\beta$ is a polynomial of degree at most $k$, of
the same parity as $k$; the monomials with $|\beta|+2\gamma\ge6$ are absorbed into
$\widetilde R_s$. Since $\Lambda-1$ is even in $x$ and vanishes at $(t;0,0)$, it is
$O(|x|^2+|y|)$, so $P_0=1$ and $P_1=0$.

\emph{Step 4 (the mean).} Let $A_s=\{|F|\le r'/s\}$, with $r'$ such that $(sf,s^2)\in U$ for
$f\in A_s$ and all small $s$; then $\P(A_s^c)\le Ce^{-c/s^2}$. At a positive root,
$\alpha\pi_0G_0(\tau)/\tau\le\alpha G(\tau)/\tau=1$, so $0\le\FDP_\infty\le1$. Hence, by
\eqref{eq:taylor},
\[
\frac{\FDR_\infty(s)}{\alpha\pi_0}=\sum_{k=0}^5s^k\,\E[P_k(F)\mathbf 1_{A_s}]+O(s^6)+O(e^{-c/s^2}).
\]
$A_s$ is symmetric, so the terms with odd $k$ vanish. For even $k$,
$\E[P_k(F)\mathbf 1_{A_s}]=\E P_k(F)+O(e^{-c/s^2})$. Therefore
$\FDR_\infty(s)/(\alpha\pi_0)=1+s^2\E P_2(F)+s^4\E P_4(F)+O(s^6)$.

\emph{Step 5 (the coefficients).} For fixed $(c,\theta,b,f)$ and $s|b|<1$, the identity
$\E[\He_l(\rho X+\sqrt{1-\rho^2}Z)\mid X]=\rho^l\He_l(X)$ gives the absolutely convergent series
$Q(c;\theta+sb^\top f,1-s^2|b|^2)=Q(c;\theta,1)+\sum_{l\ge1}\frac{\psi_l(c;\theta)}{l!}\,s^l|b|^l\He_l(b^\top
f/|b|)$. Its $l$-th term is homogeneous of degree $l$ in $s$, so the series is the Taylor series in
$s$. Integrating, and writing $q_A(f)=f^\top Af-\tr A$,
\begin{align*}
L(t;f)&=1+\tfrac{s^2}{2}\,\ell_2(t)\,q_{M_0}(f)+s^4\,\tfrac{\ell_4(t)}{24}\,w_4(f)+O(s^6),\\
G(t;f)&=G(t)+s\,a(t)^\top f+\tfrac{s^2}{2}\,q_{N(t)}(f)+O(s^3),
\end{align*}
where $w_4(f)=\int|b|^4\He_4(b^\top f/|b|)\,\nu_0(db)$ has $\E w_4(F)=0$. Write
$\tau(f)=\tstar+s\tau_1+s^2\tau_2+O(s^3)$. Matching powers of $s$ in
$D(\tau(f))+\alpha s\,a(\tau(f))^\top f+\tfrac{\alpha s^2}{2}q_N(f)+O(s^3)=0$, and using
$D'(\tstar)=-\alpha/\kappa$ and $D''=\alpha G''$, gives
\[
\tau_1=\kappa\,a^\top f,\qquad
\tau_2=\kappa\Big[\tfrac12G''\tau_1^2+(a'^\top f)\,\tau_1+\tfrac12q_N(f)\Big].
\]
Substituting into $L(\tau(f);f)$,
\[
P_2=\tfrac12\ell_2\,q_{M_0},\qquad
P_4=\tfrac12q_{M_0}\big(\ell_2'\tau_2+\tfrac12\ell_2''\tau_1^2\big)+\tfrac{\ell_4}{24}\,w_4,
\]
with all functions evaluated at $\tstar$. For $F\sim N(0,I)$, $\E q_A(F)=0$,
$\E[q_A(F)q_B(F)]=2\tr(AB)$ and $\E[q_A(F)(c^\top F)(d^\top F)]=2c^\top Ad$. Hence $\E P_2=0$, and
\[
\E P_4=\tfrac12\Big[\kappa\ell_2'\big(G''\kappa^2\,a^\top M_0a+2\kappa\,a^\top M_0a'+\tr(M_0N)\big)
+\ell_2''\kappa^2\,a^\top M_0a\Big],
\]
which is the bracket in \eqref{eq:expansion}.

\emph{Step 6 (the standard deviation).} By Steps 3 and 5, on $A_s$ we have
$\Phi(sF,s^2)=1+s^2P_2+s^3P_3+O(s^4(1+|F|)^6)$. The product $P_2P_3$ is odd, so
$\Var\Phi(sF,s^2)=s^4\Var P_2(F)+O(s^6)$. Moreover $\Var P_2(F)=\frac14\ell_2^2\cdot2\tr(M_0^2)$ and
$\rhob=s^4\tr(M_0^2)$, which give \eqref{eq:sd}.

\emph{Step 7 (finite $p$).} For $f\in A_s$, Step 2 gives (A3) at $f$, so by the proof of
Theorem~\ref{thm:limit}, $\FDP\to\FDP_\infty(f)$ almost surely given $F=f$. Dominated convergence
gives
$\E[\FDP\,\mathbf 1_{A_s}]\to\E[\FDP_\infty(F)\mathbf 1_{A_s}]$, while
$0\le\E[\FDP\,\mathbf 1_{A_s^c}]\le\P(A_s^c)$. Hence $\liminf\FDR$ and $\limsup\FDR$ lie within
$\P(A_s^c)=O(e^{-c/s^2})$ of $\FDR_\infty(s)$.

\paragraph{Corollary~\ref{cor:sign}.}
Under exchangeability, $a(t)=\pi_1\big(\int\psi_1(c_t;\theta)g_1(d\theta)\big)\int b\,\nu_0(db)$. This
vanishes if $\int b\,\nu_0=0$, or if $g_1$ is symmetric, because $\psi_1(c;-\theta)=-\psi_1(c;\theta)$.
The form of $N$ is immediate, and $\ell_2'<0$ by Lemma~\ref{lem:shape}. For (i) and (ii), write
$\psi_2(c;\theta)=k(c-\theta)+k(c+\theta)$ with $k(x)=x\phi(x)\ge-\phi(1)$. If $|\theta|\le c$, both
arguments are nonnegative and $\psi_2\ge0$, which proves (i). If $|\theta|>c$, then
$\psi_2\ge-\phi(1)$ and $Q(c;\theta,1)\ge\frac12$. Hence
$\pi_1\int\psi_2\,dg_1\ge-2\phi(1)\pi_1G_1(\tstar)=-2\phi(1)\tstar(1/\alpha-\pi_0)$, using the fixed point
$\tstar=\alpha(\pi_0\tstar+\pi_1G_1(\tstar))$. Since $\pi_0\psi_2(\cstar;0)=\pi_0\tstar\cstar\lambda(\cstar)$,
we get $\Psi_2(\tstar)>0$ under (ii).


\begin{thebibliography}{99}\small
\bibitem[Benjamini and Hochberg(1995)]{bh1995} Y.~Benjamini and Y.~Hochberg. Controlling the false discovery rate: a practical and powerful approach to multiple testing. \emph{J. R. Stat. Soc. B}, 57:289--300, 1995.
\bibitem[Benjamini and Yekutieli(2001)]{by2001} Y.~Benjamini and D.~Yekutieli. The control of the false discovery rate in multiple testing under dependency. \emph{Ann. Statist.}, 29:1165--1188, 2001.
\bibitem[Benjamini(2010)]{benjamini2010} Y.~Benjamini. Discovering the false discovery rate. \emph{J. R. Stat. Soc. B}, 72:405--416, 2010.
\bibitem[Clarke and Hall(2009)]{clarke2009} S.~Clarke and P.~Hall. Robustness of multiple testing procedures against dependence. \emph{Ann. Statist.}, 37:332--358, 2009.
\bibitem[Delattre and Roquain(2011)]{delattre2011} S.~Delattre and E.~Roquain. On the false discovery proportion convergence under Gaussian equi-correlation. \emph{Statist. Probab. Lett.}, 81:111--115, 2011.
\bibitem[Dobriban(2026)]{dobriban2026} E.~Dobriban. The Benjamini--Hochberg procedure can fail to control the FDR for correlated two-sided Gaussian tests. \emph{Proc. Natl. Acad. Sci. USA}, 123(41):e2626267123, 2026. arXiv:2607.12208.
\bibitem[Efron(2007)]{efron2007} B.~Efron. Correlation and large-scale simultaneous significance testing. \emph{J. Amer. Statist. Assoc.}, 102:93--103, 2007.
\bibitem[Fan et~al.(2012)]{fan2012} J.~Fan, X.~Han and W.~Gu. Estimating false discovery proportion under arbitrary covariance dependence. \emph{J. Amer. Statist. Assoc.}, 107:1019--1035, 2012.
\bibitem[Ferreira and Zwinderman(2006)]{ferreira2006} J.~A. Ferreira and A.~H. Zwinderman. On the Benjamini--Hochberg method. \emph{Ann. Statist.}, 34:1827--1849, 2006.
\bibitem[Finner et~al.(2007)]{finner2007} H.~Finner, T.~Dickhaus and M.~Roters. Dependency and false discovery rate: asymptotics. \emph{Ann. Statist.}, 35:1432--1455, 2007.
\bibitem[Ghosh and Sarkar(2026)]{ghosh2026} D.~Ghosh and S.~K. Sarkar. Further results on controlling the false discovery rate in two-sided Gaussian mean testing. arXiv:2608.21267, 2026.
\bibitem[GTEx Consortium(2017)]{gtex2017} GTEx Consortium. Genetic effects on gene expression across human tissues. \emph{Nature}, 550:204--213, 2017.
\bibitem[Genovese and Wasserman(2004)]{genovese2004} C.~Genovese and L.~Wasserman. A stochastic process approach to false discovery control. \emph{Ann. Statist.}, 32:1035--1061, 2004.
\bibitem[Guo and Rao(2008)]{guo2008} W.~Guo and M.~B. Rao. On control of the false discovery rate under no assumption of dependency. \emph{J. Statist. Plann. Inference}, 138:3176--3188, 2008.
\bibitem[Lei(2026)]{lei2026} L.~Lei. How much can Gaussian dependence inflate the Benjamini--Hochberg procedure's FDR? arXiv:2607.14812, 2026.
\bibitem[Reiner-Benaim(2007)]{reiner2007} A.~Reiner-Benaim. FDR control by the BH procedure for two-sided correlated tests with implications to gene expression data analysis. \emph{Biom. J.}, 49:107--126, 2007.
\bibitem[Sarkar(2002)]{sarkar2002} S.~K. Sarkar. Some results on false discovery rate in stepwise multiple testing procedures. \emph{Ann. Statist.}, 30:239--257, 2002.
\bibitem[Schwartzman and Lin(2011)]{schwartzman2011} A.~Schwartzman and X.~Lin. The effect of correlation in false discovery rate estimation. \emph{Biometrika}, 98:199--214, 2011.
\bibitem[Shang and Guo(2026)]{shang2026} Z.~Shang and W.~Guo. A dimension-free bound on the FDR of the Benjamini--Hochberg procedure under slowly decaying Gaussian correlations. arXiv:2611.32898, 2026.
\bibitem[Storey et~al.(2004)]{storey2004} J.~D. Storey, J.~E. Taylor and D.~Siegmund. Strong control, conservative point estimation and simultaneous conservative consistency of false discovery rates: a unified approach. \emph{J. R. Stat. Soc. B}, 66:187--205, 2004.
\bibitem[Sun and Stephens(2018)]{sun2018} L.~Sun and M.~Stephens. Solving the empirical Bayes normal means problem with correlated noise. arXiv:1812.07488, 2018.
\bibitem[Zhang(2026)]{zhang2026} X.~Zhang. Exact Simes inequalities for two-sided Gaussian tests. arXiv:2611.02513, 2026.
\end{thebibliography}
\end{document}